\documentclass[a4paper,12pt]{iopart}
\usepackage{amstext}
\usepackage[utf8]{inputenc}

\usepackage{graphicx}
\usepackage{url}
\usepackage{upgreek}
\newcommand{\uK}{$\upmu$K}
\newcommand{\um}{$\upmu$m}
\newcommand{\us}{$\upmu$s}

\begin{document}

\title{Single-atom trapping and transport in DMD-controlled optical tweezers}

\author{Dustin Stuart and Axel Kuhn}
\address{Clarendon Laboratory, Parks Road, Oxford, OX1 3PU}
\ead{axel.kuhn@physics.ox.ac.uk}

\date{\today}
\begin{abstract}
We demonstrate the trapping and manipulation of single neutral atoms in reconfigurable arrays of optical tweezers. Our approach offers unparalleled speed by using a Texas Instruments Digital Micro-mirror Device (DMD) as a holographic amplitude modulator with a frame rate of 20,000 per second. We show the trapping of static arrays of up to 20 atoms, as well as transport of individually selected atoms over a distance of 25\,\um\ with laser cooling and 4\,\um\ without. We discuss the limitations of the technique and the scope for technical improvements.
\end{abstract}
\maketitle

\section{Introduction}

Single neutral atoms are promising candidates for physically realising quantum bits (qubits), the fundamental unit of quantum information. An impressive list of schemes for performing quantum logic gates have been demonstrated using controlled collisions \cite{Kaufman:2015aa, Dorner:2005kr} and Rydberg states \cite{Leseleuc:2017aa, Jau:2016aa, Xia:2015aa, Wilk:2010bs}. The current challenge is to scale these demonstrations to larger numbers of qubits, which demands a method for trapping large, reconfigurable arrays of independently addressable atoms.

The most common traps for neutral atoms are those based on the optical dipole force \cite{Corwin:1999km, Schlosser:2002bu}. Large arrays of dipole traps have been produced using optical lattices \cite{Nelson:2007ct} and microlens arrays \cite{Schlosser:2012aa, Lengwenus:2010yb}, however, in all these systems, the trapping sites can only be moved in unison, not individually. Several approaches have been tried for creating reconfigurable traps, for example, acousto-optic deflectors (AODs) \cite{Barredo:2016aa, Endres:2016aa, Boyer:2004mq} and liquid crystal spatial light modulators (SLMs) \cite{Kim:2016aa, Nogrette:2014pt, Bowman:2017aa}. AODs are fast but can only move traps in one dimension at a time. SLMs allow full 2D manipulation but they are limited to frame rates of 60\,Hz due to the relaxation time of the liquid crystal and thus too slow for many prospective applications.

In this paper, we overcome this limitation by using a Digital Mirror Device (DMD) \cite{Muldoon:2012ud, :2005aa} to holographically generate arrays of independently movable dipole traps. Our DMD (Texas Instruments DLP Discovery 1100) is a $1024 \times 768$ array of micro-mechanical mirrors. Each mirror can be switched between two angles ($-12^\circ$ and $+12^\circ$), which we refer to as `on' and `off'. With a full frame rate of 20\,kHz our DMD is much faster than any liquid crystal SLM. Furthermore, using the DMD in the Fourier plane also allows for the correction of optical aberrations in the experimental apparatus, which is essential for producing tightly focussed dipole traps.

This paper is divided into three sections. Firstly, we describe the experimental setup for cooling and trapping atoms, and characterise the properties of a single dipole trap. Secondly, we show how the DMD can be used to generate arbitrary arrangements of up to 20 single atoms. Finally, we demonstrate simultaneous transport of single atoms to arbitrary locations and discuss the capabilities and limitations of the technique.

\section{Trapping and imaging single atoms}

\begin{figure}
\includegraphics[width=\columnwidth]{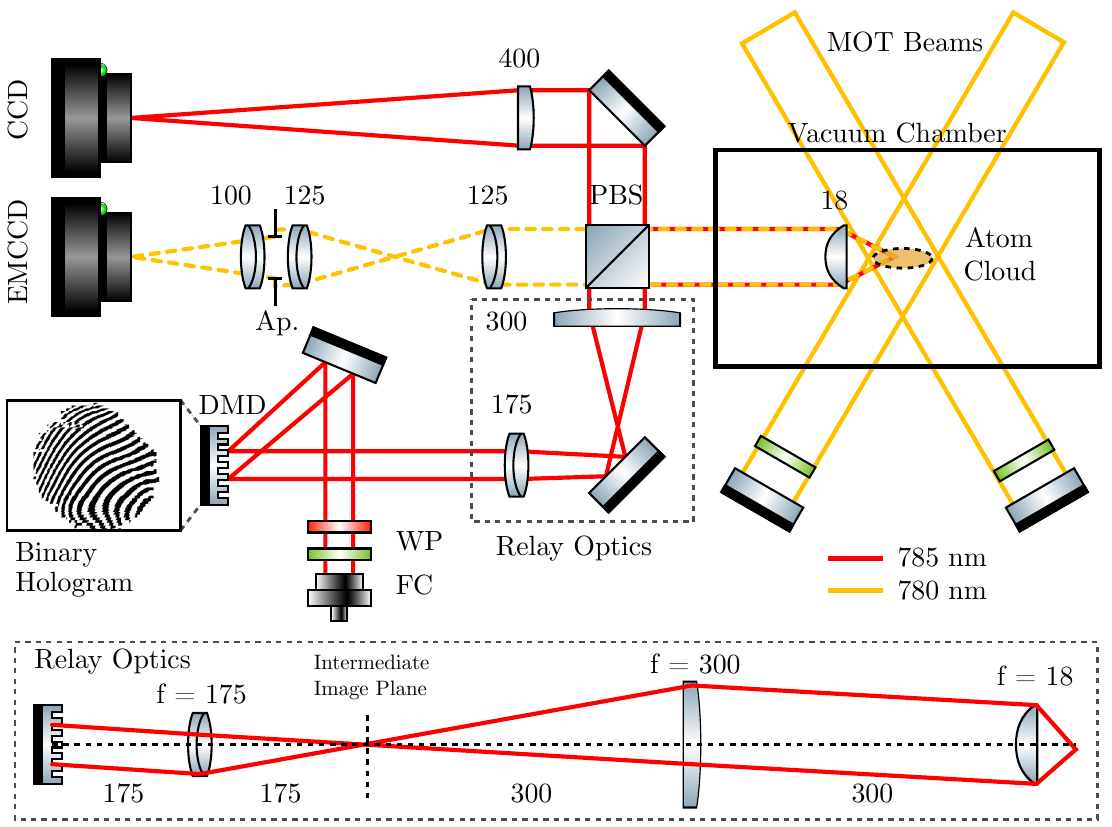}
\caption{Diagram of the experimental setup (all dimensions in millimetres). A magneto-optical trap (MOT) formed from three retro-reflected cooling laser beams confines a cloud of Rubidium atoms at the centre of a vacuum chamber. Trapping light at 785\,nm from the tapered amplifier is delivered via a single mode fibre collimator (FC) on to the DMD. The DMD imprints a binary amplitude hologram of the desired arrangement of traps on to the beam, which is transferred via relay optics on to an aspheric lens ($f = 18$\,mm) inside the vacuum chamber. The microscopic dipole traps are formed in the focal plane of this lens. A polarising beam splitter (PBS) reflects most of the trapping light into the vacuum chamber, and transmits the remainder to a CCD camera for monitoring the arrangement of the traps. The relay optics are arranged to form a \emph{ff-f'f'} telescope, which ensures that light rays from the DMD will always arrive at the aspheric lens regardless of their angle. Fluorescence light at 780\,nm from single trapped atoms is collected by the same aspheric lens and imaged on to an EMCCD camera. An aperture (Ap.) eliminates stray light from the MOT beams. Created using ComponentLibrary by Alexander Franzen \url{http://www.gwoptics.org/ComponentLibrary/}}
\label{setup}
\end{figure}

Our experimental apparatus (see fig. \ref{setup}) is essentially the combination of a magneto-optical trap (MOT) \cite{Raab:1987aa} and a high numerical aperture microscope. The MOT acts as a reservoir of $\sim 10^4$ laser-cooled {Rubidium-87} atoms with a temperature of $26$\,\uK. The cooling lasers are tuned to the $F = 2 \rightarrow F^\prime = 3$ hyperfine component of the 5s~$^2$S$_{1/2} \rightarrow$ 5p~$^2$P$_{3/2}$ transition at 780\,nm. At the centre of the MOT, we impose additional microscopic dipole traps. The traps are formed from a far off-resonant laser beam from a 1\,W tapered amplifier laser (Toptica {DLX 110}, $\lambda = 785.4$\,nm) which is focussed by an aspheric lens (Asphericon {HPX 20-18}) with a numerical aperture NA~$= 0.45$. This laser induces a light shift of the ground state to lower energy by an amount

\begin{equation}
U_0= \frac{\hbar \Gamma^2}{24} \frac{I}{I_{\text{sat}}} \left(\frac{1}{\Delta_{1/2}} + \frac{2}{\Delta_{3/2}}\right)
\end{equation}

\noindent where $\Gamma = 2\pi\times6$\,MHz is the natural linewidth of the transition, $I_\text{sat} = 2\hbar\pi^2c\Gamma/3\lambda^3 = 16.7$\,Wm$^{-2}$ is the saturation intensity, and $\Delta_{1/2}, \Delta_{3/2}$ are the frequency detunings from the 5s~$^2$S$_{1/2}\rightarrow$~5p~$^2$P$_{1/2,3/2}$ transitions respectively. We model the intensity of the laser and hence the dipole potential as a Gaussian beam $U_\text{dip}(r) = U_0\exp(-2r^2/w_0^2)$. For a laser power of 1.0\,mW and a waist $w_0 = 1.4$\,\um\ we expect a trap depth $U_0 = -k_B \times 0.4$\,mK.

The MOT beams both cool the atoms into the dipole trap and scatter photons which allow the atoms to be detected. For loading the dipole traps, we set the total intensity of the MOT beams to $I = 140$\,Wm$^{-2}$ and the detuning to $\Delta = -5.7\Gamma$. The scattered light is collected using the same high-NA lens and imaged on to an EMCCD camera (Andor iXon 885). The magnification of the imaging system was chosen so that a diffraction-limited point source is focussed on to a single camera pixel. The aperture of the imaging system was limited to $d = 15.2$\,mm so as to eliminate unwanted stray light from the MOT beams that is reflected from the edges of the lens.

When calculating the overall detuning of the cooling laser, we must include the light shifts of the ground and excited states. The dipole trapping light is linearly polarised so the ground state shift $\Delta_\text{ls} = U_0/\hbar = 2\pi \times 8.4$\,MHz is the same for all $m_F$ sublevels (where $m_F$ is defined with respect to the polarisation axis). Unfortunately, the excited state has a complicated tensor light shift \cite{Mathur:1968aa} which depends on $|m_F|$, for example, the $m_F = 0$ sublevel has an (upward) shift of $-\frac{3}{40}I\hbar\Gamma^2/I_\text{sat}\Delta$ while the $m_F = \pm3$ sublevels have a shift of zero. Furthermore, the upper state has a small (downward) light shift due to higher lying states such as 5d~$^2$D$_{3/2,5/2}$ at 776\,nm and 7s~$^2$S$_{1/2}$ at 741\,nm. When estimating the scattering rate, we assume an upper state light shift of zero and an effective saturation intensity of $I_\text{sat}^\prime = \frac{15}{7}I_\text{sat} = 35.8$\,Wm$^{-2}$ (by averaging over the Clebsch-Gordan coefficients of all possible $F=2 \rightarrow F^\prime=3$ transitions). The scattering rate is then calculated using the equation

\begin{equation}
R_\text{sc} = \frac{\Gamma}{2} \frac{I/I_\text{sat}^\prime}{1 +
4(\Delta/\Gamma)^2 + I/I_\text{sat}^\prime}
\end{equation}

\noindent to be 360\,kHz. The overall collection efficiency was 0.6\%, which is due to the fraction of the solid angle collected by the lens (5.3\%), random splitting of the light at the PBS (50\%), the additional aperture (71\%) the quantum efficiency of the EMCCD camera (40\%) and imperfect transmission of the optics (81\%), giving an expected count rate of 2.2\,kHz.

\begin{figure}
\includegraphics{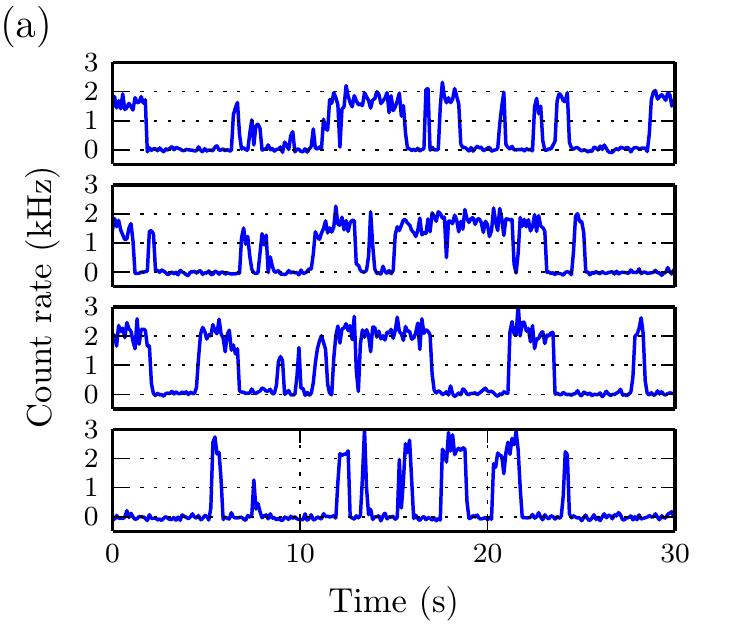}
\includegraphics{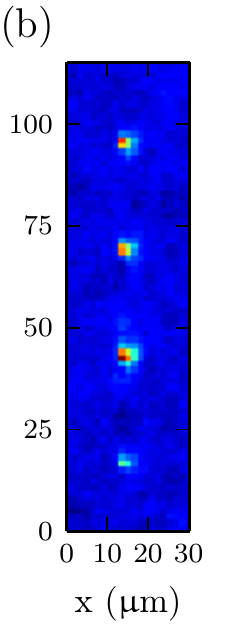}
\includegraphics{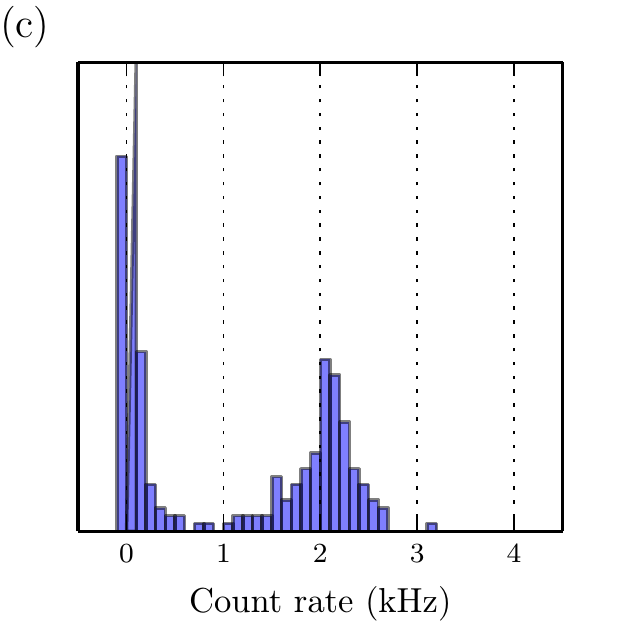}
\caption{Statistical analysis of the light emitted from trapped atoms. The background (composed of stray light and camera dark counts) is subtracted from the images. (a) Photon count rate for each dipole trap. The intervals of high count rate indicate the presence of a single atom. (b) Image of the 4 traps averaged over 30 seconds. The average size of the atom cloud is $\sigma = 1.2$\,\um. (c) Histogram of the photon counts showing two clearly resolved peaks (due to zero and one atom). The average count rate due to a single atom is 2.0\,kHz. By setting a threshold halfway between the two peaks, the probability of observing no atom is $p_0 = 0.59 \pm 0.03$, one atom is $p_1 = 0.40 \pm 0.03$, and two atoms is $p_2 < 0.01$.}
\label{four}
\end{figure}

Figure \ref{four} shows the counts from four individual dipole traps. The number of counts jumps between two distinct values indicating the presence or absence of a single atom. The measured count rate for a single atom was 2.0\,kHz, which agrees well with the expected value. The size of the atom cloud from a Gaussian fit yields $\sigma = 1.2$\,\um, which is somewhat larger than the diffraction limit of the imaging system ($\sigma = 0.4$\,\um). The larger size is due to the fact that the longitudinal extent of the dipole trap is $z_R = \pi w_0^2/\lambda = 5.6$\,\um, so the closest and farthest atoms will be viewed out of focus. The number of atoms in the trap is limited to either zero or one by the collisional blockade mechanism \cite{Schlosser:2002bu, Fung:2016aa}: when pairs of Rubidium atoms collide in the presence of red-detuned cooling light, they can be photo-excited to form a molecule and are lost from the trap. This is confirmed in figure \ref{four}c, which shows a negligible probability of observing two atoms. The expected one-atom occupation probability is $p_1 = 0.5$ in the collisional blockade regime (although this can be increased to $p_1 > 0.8$ using blue-detuned light \cite{Fung:2015aa}). We measure $p_1 = 0.40$, which implies that there is another mechanism for atom loss in addition to the collisional blockade.

\begin{figure}
\includegraphics{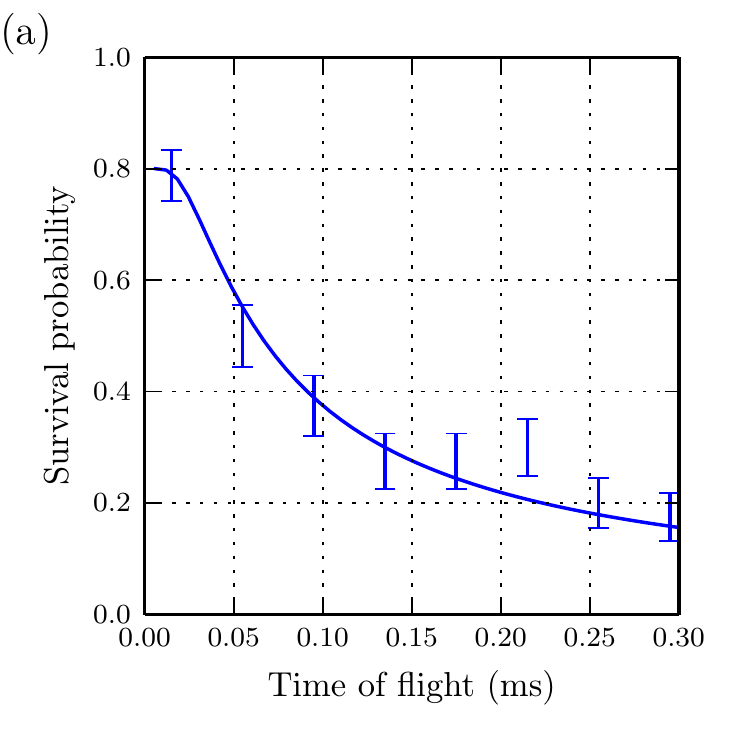}
\includegraphics{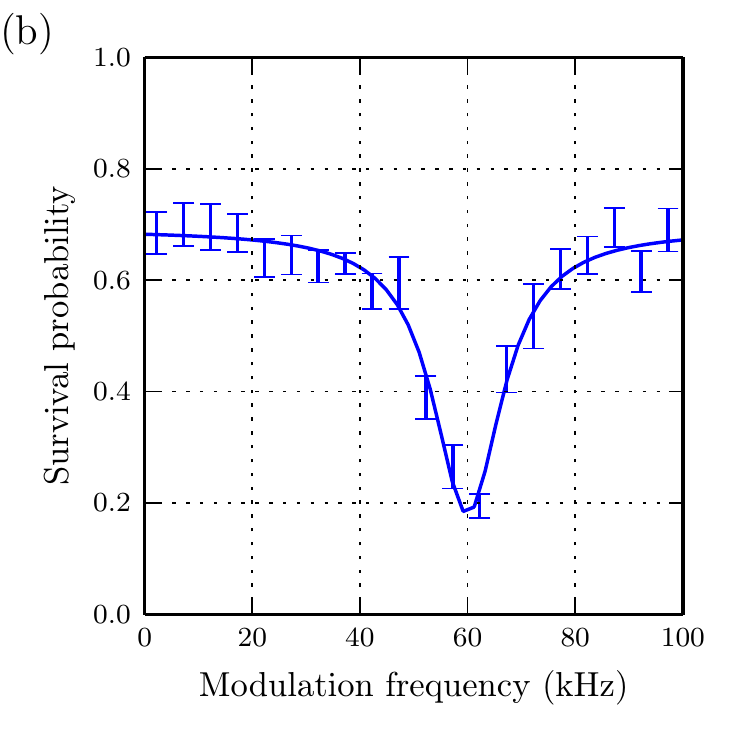}
\caption{(a) Measurement of temperature by release and recapture. The temperature from the fitted model is $16\pm2$\,\uK, which is typical of
sub-Doppler cooled atoms. (b) Measurement of the trap oscillation frequencies by modulating the trap depth. The radial oscillation frequency was determined from a Lorentzian fit to be $\omega_\text{r} = 2\pi\times30.0 \pm 0.2$\,kHz.}
\label{parametric}
\end{figure}

We measured the lifetime of single atoms in the dipole trap by switching off the cooling lasers and measuring the survival probability as a function of time. The $1/e$ lifetime was found to be $\tau = 1.4$\,s. This is much shorter than the average time between collisions from room-temperature background gas atoms, which was determined from the lifetime of the MOT to be $\tau = 24 \pm 1$\,s. Rather, the measured lifetime of the dipole trapped atoms is consistent with the lifetime that would be expected due to heating from the dipole trapping laser itself. The scattering rate due to the dipole trapping laser is $R_\text{sc} = 160$\,s$^{-1}$, with each scattering event heating the atom by $E_\text{r} = h^2/m\lambda^2 = k_B\times3.6$\,nK. From this, we estimate the theoretical lifetime to be $U_\text{dip}/E_\text{r}R_\text{sc} = 5.6$\,s (starting from an atom in the motional ground state).

We determined the temperature of the atoms in the dipole trap by a release-and-recapture method \cite{Tuchendler:2008aa, Alt:2003aa}. We assume that atoms are at the centre of the dipole trap and that the radial velocity $v$ follows a 2D Maxwell-Boltzmann distribution $p(v) = \frac{mv}{k_\text{B}T}\exp(-mv^2/k_\text{B}T)$. We switch off the trap for a time $t$ during which the atom travels a distance $r = vt$. When the trap is switched back on, the atoms have gained potential energy. At a certain critical velocity $v_\text{c}$, the atoms will have enough energy to escape the trap, i.e. when $\frac12mv_\text{c}^2 - U_0\exp(-2v_\text{c}^2t^2/w_0^2) > 0$. The probability of recapturing the atom is given by $p = \int_0^{v_c}p(v)dv$. We determined the temperature of the atoms to be $16\pm2$\,\uK\ by fitting this model to the measured recapture probability (see fig. \ref{parametric}a).

Finally, we experimentally determined the trap frequency by parametric loss spectroscopy (see fig. \ref{parametric}b). We modulated the intensity of the trapping laser by 20\% for 100\,ms over a frequency range of $0 < f_\text{mod} < 100$\,kHz. When the modulation frequency is equal to twice the oscillation frequency, atoms are parametrically driven out of the trap. The expected value of the radial oscillation frequency $\omega_\text{r} =
\sqrt{4U_0/m w_0^2} = 2\pi\times35$\,kHz which agrees well with the fitted value of $\omega_\text{r} = 2\pi\times30.0 \pm 0.2$\,kHz. The expected longitudinal oscillation frequency $\omega_z = \sqrt{2U_0/m z_R^2} = 2\pi\times3.9$\,kHz was too slow to be observed.

\section{Trapping arrays of atoms}

We now demonstrate that our setup can be scaled to large numbers of traps. We use the DMD to imprint a binary amplitude mask on to the trapping beam. The desired arrangement of traps (the Fourier transform of the mask) is produced in the focal plane of the aspheric lens.

As simple example, we consider a single focussed trap. The complex-valued field of this trap in the plane of the DMD is given by
\begin{equation}
E(x,y) = A\exp(i\frac{2\pi}{f\lambda}(x^\prime_0x + y^\prime_0y))
\end{equation}
\noindent where $f = 30.9$\,mm is the effective focal length of the lens (including the relay optics), $x,y$ are the coordinates on the DMD and $x^\prime_0, y^\prime_0$ are the coordinates of the trap. The easiest way to convert this complex field to a binary amplitude hologram is to switch only those mirrors `on' which satisfy $0 < \arg(E) <\pi$, leading to an average 50\% filling ratio. The hologram for a single trap consists of stripes of mirrors which alternate between `on' and `off', forming an artificial diffraction grating. The first-order diffraction peak can be moved by changing the period and angular orientation of the grating, allowing for arbitrary positioning of the trap. The power in the $n^{th}$ order of diffraction is given by $\frac14\text{sinc}^2(n\pi/2)$ \cite{Goodman:1996gs}, which for the first order is 10.1\,\%.

The case for multiple traps is more complicated. Ideally, one would simply sum up the complex-valued holograms for each trap. However, the act of converting this to a binary amplitude hologram results in drastic variation of the trap depths, as well as generating unwanted ghost traps. These problems can be avoided by dithering the hologram, by using an iterative phase retrieval algorithm such as Gerchberg-Saxton \cite{Gerchberg:1972zp}, or by conjugate gradient minimisation techniques \cite{Bowman:2017aa}. In a separate publication, we carefully consider the merits of these different algorithms \cite{holodmd}. For the case of a static array of traps, we use the weighted Gerchberg-Saxton algorithm, which is slow but allows for the generation of very accurate trapping potentials.

To mimic other optical elements or to compensate for aberrations, the initial complex-valued field can be multiplied by an arbitrary phase map $\exp(i\phi(x,y))$. We take advantage of this feature for two purposes: firstly, we apply a quadratic phase function $\phi(x,y) = \frac{2\pi z^\prime_0}{2f^2\lambda}(x^2+y^2)$, which acts as an artificial lens allowing us to adjust the longitudinal position of the traps by $z^\prime_0$. Secondly, we add a custom $\phi(x,y)$ to correct for wavefront errors caused by the subsequent optical elements. The largest contribution to the wavefront error (10\,$\lambda$) comes from the DMD itself, resulting in a spot size that is much larger than the diffraction limit. We measure these wavefront errors \emph{in situ} by using the CCD camera to observe interference patterns between different regions of the DMD and use this information to calculate $\phi(x,y)$ (see \cite{holodmd} for details). This allows us to correct for aberrations along the entire optical path.

\begin{figure*}
\includegraphics{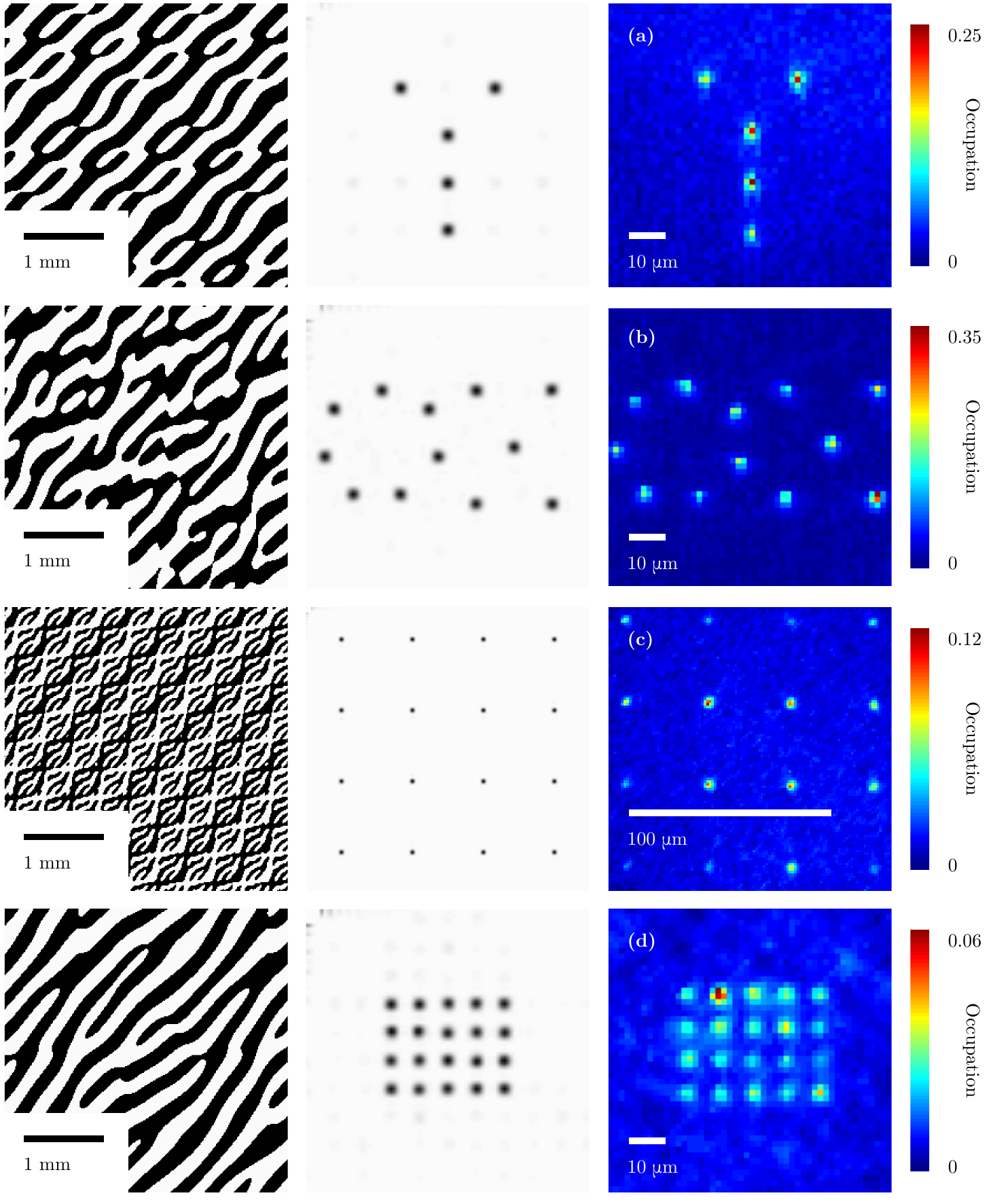}
\caption{Gallery of trapped atoms with various arrangements. For each case, we show a portion of the binary amplitude hologram, simulated trapping potential, and fluorescence images of the trapped atoms. These images are averaged for 30 seconds, excluding frames where $< 2$ of the traps are filled. (a) the letter `Y' with 5 traps (mean trap occupation probability $p = 0.22$), (b) the letters `OX' with 12 traps ($p = 0.12$), (c) a widely spaced grid of 16 traps ($p = 0.1$) and (d) a grid of 20 traps ($p = 0.044$). The traps are not equally occupied for several reasons: the position of the MOT is unstable such that its centre fluctuates randomly within the trapping region (b); the field-of-view of the aspheric lens is limited so that traps further from the centre are not as deep (c); and, while the theoretical trap depths are even, the actual trap depths are not because of aberrations in the optical setup (d) \cite{Nogrette:2014pt}.}
\label{atoms}
\end{figure*}

Our holographic technique is capable of generating arbitrary configurations of trapping sites, some examples of which are shown in figure \ref{atoms}. The maximum displacement of a single trap is 215\,\um, which is determined by the maximum spatial frequency of the hologram on the DMD. In practice, the small field of view of the microscope objective impose a smaller limit.

In theory, the maximum number of resolvable trapping sites is determined by the number of pixels on the DMD. A more stringent limit is imposed by the Gerchberg-Saxton algorithm, which performs poorly with large numbers of trapping sites. Furthermore, our laser power is reduced due to about 200\,mW due to fibre-coupling efficiency (40\%), the finite reflectivity of the DMD (68\%) \cite{:2005aa}, and other optics (80\%). Using this modest laser power, we found we could generate a maximum of 20 trapping sites. Titanium sapphire lasers are available with 20\,W, which would yield a 100-fold increase in the number of traps. The average occupation probability is limited to $p < 0.5$ in the collisional blockade regime. We observe occupation probabilities less than this because of the additional loss due to scattering from the dipole trap.
  
\section{Transporting single atoms}

\begin{figure}
\includegraphics{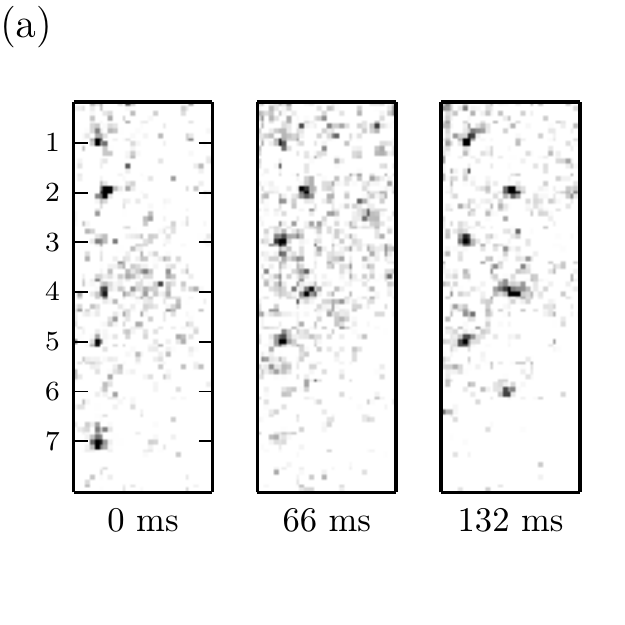}
\includegraphics{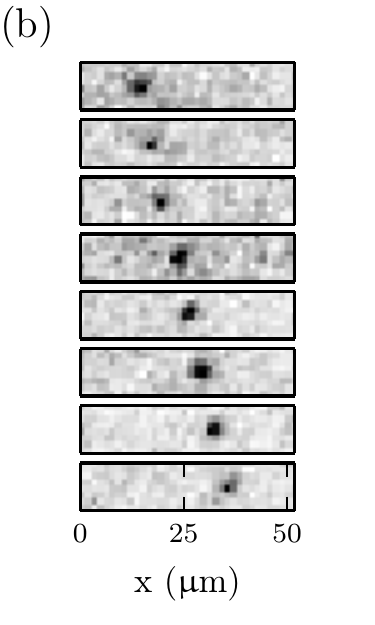}
\includegraphics{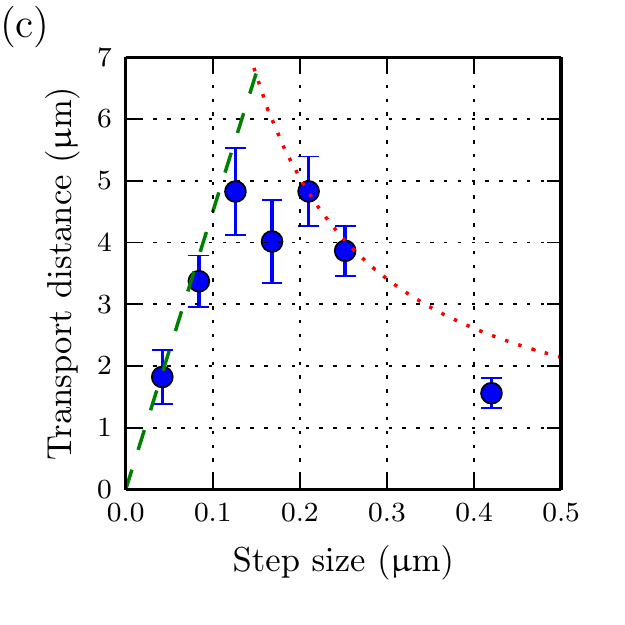}
\caption{Transport of atoms. (a) Several frames from a movie showing independent transport of an array of 7 atoms. Traps 1, 3, 5 and 7 are stationary while traps 2, 4 and 6 are moved in 30 steps of 0.89\,\um, giving a total distance of 25\,\um. As in fig. \ref{four}, atoms randomly enter and exit the trap while undergoing continuous laser cooling. (b) A sequence of 8 frames from the EMCCD camera showing a single atom being transported in trap 4. (c) Transport of atoms without laser cooling. The transport distance is defined to be the distance at which the probability of atom loss is $1/e^2$. For small step sizes, the maximum transport distance is limited by heating due to the mirror settling time of the DMDs (green dashed line). For large step sizes, the maximum distance is limited due to the high probability of atom loss at each step (red dotted line).}
\label{transport}
\end{figure}

We now demonstrate the transport of atoms by moving the trapping site. To do this, we simply display a sequence of holograms on the DMD in which the trapping site is moved in small steps $\delta x$. Figure \ref{transport} illustrates the transport of 7 single atoms in this way, while being continuously cooled. The cooling suppresses any heating effects experienced by the atom, allowing us to transport atoms over a large distance of 25~\um . By performing the transport with a small enough step size, we can adiabatically move an atom between two positions. We can estimate the heating rate due to the finite step size as follows: assuming an atom is at rest in the centre of the trap, the energy is increased by $E_s = m\omega_\text{r}^2\delta x^2/2$ for each step.

The DMD introduces an additional source of heating. When the micro-mirrors are switched between the `on' and `off' states, they have a finite settling time of 12~\us . During this time, the mirror vibrates by several degrees about it's equilibrium angle of 12$^\circ$. The angular displacement during vibration is such that the light no longer passes through the relay optics, which causes the intensity of the trap to flicker. We can estimate the energy gained by the atom as $E_\text{m} = \frac12m\omega_\text{r}^2v^2t^2 = 44$~\uK. This source of heating is purely technical, and could be overcome by using lenses with larger diameters in the relay optics, or by mechanically damping the DMD micro-mirrors. We measured the severity of this heating by observing the reduction in lifetime for different frame rates of the DMD while the trap was held stationary. At a frame rate of 50~s$^{-1}$, the lifetime is reduced by a factor of 2 indicating that the heating rate due to mirror switching is equal to that of photon scattering.

Finally, we demonstrate dark transport of a single atom. The atoms were transported for a constant time of 0.5~s at a frame rate of 50~s$^{-1}$. The step size $\delta x$ determines the dominant source of heating and hence the maximum transport distance (see fig. \ref{transport}). For small step sizes, the transport distance is limited by intensity flicker due to the mirror settling time. For large step sizes the dominant source of heating is that the trap does not move smoothly. The optimum step size that balances these two effects was found to be 0.2~\um, yielding a maximum transport distance of 4~\um.

\section{Conclusion}

We have demonstrated cooling, trapping and transport of atoms with holographically generated optical tweezers using a digital mirror device. We generate a wide variety of possible trap configurations, which allows one to design static arrays of atoms with flexible nearest-neighbour connectivity; for transporting and positioning a single atom in the mode of a high-finesse cavity; or for trapping a reservoir of atoms in a high-density region (e.g. a MOT) for later transport into a low-density region for quantum operations.

For stationary traps, the lifetime of the atoms is limited to 1.4\,s by off-resonant scattering from the dipole trapping laser. This is much shorter than the expected limit due to background gas collisions of 24\,s. A simple solution would be to increase the detuning of the dipole trapping laser from 5\,nm to $> 50$\,nm, as has been done in other experiments \cite{Kim:2016aa, Nogrette:2014pt, Endres:2016aa}. For a constant trap depth, the rate of off-resonant scattering scales as $1/\Delta$, while the required power increases linearly with $\Delta$ however, at very large detunings (e.g. $\lambda=1064$\,nm), quantum interference effects mean that rate of spin-changing scattering events is suppressed even further \cite{Cline:1994aa}.

We demonstrate two kinds of transport: with and without laser cooling. Transport with cooling is much more robust against atom loss. In this case, the lifetime is similar to the lifetime of static traps, and is limited by off-resonant laser scattering. By switching to a far-detuned laser we would expect a dramatic increase in transport distance. Therefore, this type of transport would be ideal for moving atoms from a reservoir in to the mode of a high-finesse optical cavity.

On the other hand, transport without cooling has the advantage of preserving the internal quantum state of the atom. This is necessary for collision-based quantum gates and addressable quantum memories. For transport without cooling, distance is limited by re-settling oscillations of the DMD micro-mirrors. These limitations are purely technical and therefore significant improvements are possible with some simple modifications to the experiment such as larger relay lenses. Alternatively, one could use two DMDs to display successive frames, and smoothly switch between them by ramping down the laser power on the first while ramping up the laser power on the second. This would be similar to the 'double buffering' technique used in computer graphics to produce smooth animations.

In conclusion, our single-atom tweezers have wide-ranging applications in quantum technologies, in particular when single atoms need to be manipulated accurately, for example, close to dielectric mirrors in optical cavities. Furthermore, we have characterised the limitations of our setup, which are technical in nature, and identify potential improvements.

\ack

We acknowledge support for this work through the quantum technologies programme (NQIT hub).

\section*{References}

\bibliography{atomtrap}

\end{document}